\documentclass[%superscriptaddress,
twocolumn,
showpacs,preprintnumbers,
bibnotes,
 amsmath,amssymb,
 aps,
 pra,
superscriptaddress,
longbibliography,
]{revtex4-1}

\usepackage[version=3]{mhchem} % Formula subscripts using \ce{}
\usepackage[dvipsnames]{xcolor}
\usepackage{textcomp}
\usepackage{graphicx}
\usepackage{amsmath}
\usepackage{fixmath}
\usepackage{amsfonts}
\usepackage{amssymb}

\begin{document}

\title{Quantum spin compass models in 2D electronic topological metasurfaces}% Force line breaks with \\

\author{F.~O. Nigmatulin}
\affiliation{%
 Faculty of Physics, ITMO University, St. Petersburg 197101, Russia}%
\author{I.~A. Shelykh}
\affiliation{Science Institute, University of Iceland IS-107, Reykjavik, Iceland}
\affiliation{%
 Faculty of Physics, ITMO University, St. Petersburg 197101, Russia}%

\author{I.~V. Iorsh}%
\affiliation{%
 Faculty of Physics, ITMO University, St. Petersburg 197101, Russia}%

\date{\today}% It is always \today, today,

\begin{abstract}
We consider a metasurface consisting of a square lattice of cylindrical antidots in a two-dimensional topological insulator (2DTI). Each antidot supports a degenerate Kramer's pair of eigenstates formed by the helical topological edge states. We show that the on-site Coulomb repulsion leads to the onset of the Mott insulator phase in the system in a certain range of experimentally relevant parameters. Intrinsic strong spin-orbit coupling characteristic for the  2DTI supports a rich class of the emerging low-energy spin Hamiltonians which can be emulated in the considered system, which makes it an appealing solid state platform for quantum simulations of strongly correlated electron systems.
\end{abstract}

\maketitle

%\tableofcontents

Spin lattice models are ubiquitous in theoretical physics. Besides their natural applications for the description of the behavior of magnetic systems, a variety of the condensed matter problems, related to high-temperature superconductivity \cite{Lee2006}, thin superfluid films \cite{Nelson1979}, quantum Hall bilayers \cite{Kyriienko2015}, and nonlinear optical lattices \cite{Kalinin2020} allow mapping into spin lattice Hamiltonians. An interesting subclass of such  models is represented by Compass models (CM), for which the characteristic feature is direction-dependent spin-spin interaction~\cite{RevModPhys.87.1}. The first model of this type was introduced back in the 1982~\cite{kugel1982jahn} to describe the interplay between Jahn-Teller effect and magnetization dynamics. Since then CMs have been applied for modelling of emergent phenomena in variety of strongly correlated systems such as high-temperature superconductors~\cite{PhysRevLett.101.087004,PhysRevB.79.054504}, vacancy centers networks~\cite{PhysRevB.86.134412}, colossal magnetoresistance manganites~\cite{tokura1999colossal}, and materials supporting spin-liquid phases~\cite{baek2017evidence}. One of the most prominent examples is Kitaev model~\cite{kitaev2003fault} employed extensively in the rapidly developing field of topological quantum computation.  

For any spin model, it is highly desirable to find a material platform which allows flexible control over its effective parameters ~\cite{cubitt2018universal}. While it has been argued that certain quantum CMs can be emulated with use of cold atoms in optical lattices~\cite{wu2008orbital,radic2012exotic}, corresponding solid state platforms are still yet to appear. Here, we demonstrate that a metasurface formed by a square lattice of antidots in two-dimensional topological insulators (2DTI) is an attrective alternative for this. 

2DTIs are materials that  have both a 2D bulk energy gap (like ordinary insulators) and 1D gapless conducting edges  ~\cite{RevModPhys.82.3045,KANE20133,Bernevig_Hughes2013}, protected by time reversal symmetry and characterized by the spin-momentum locking, which means that at a specific boundary the direction of the propagation of an edge state is uniquely defined by electrons spin projection. Naturally, for the closed boundaries, the energy of an edge state is quantized, and for the case of rotational symmetry the corresponding eigenstates are characterized by specific projections of the orbital and spin angular momentum.  Such \textit{topological resonators} have been actively studied recently in topological photonic systems~\cite{PhysRevB.101.205303,jalali2020semiconductor,mehrabad2020chiral}, but can be realized as well for electrons, example being an antidot (ring shape aperture) in 2DTI. The electronic spectrum of such a system was obtained in~Refs. \cite{Michetti_2011,PhysRevB.84.035307} and associated quantum impurity models have been considered in a number of follow up works~\cite{lu2013quantum,xin2015kondo}. 

In the current paper, we show that a square lattice of antidots in 2DTI  emulates a quantum spin compass model with additional spin-orbit interaction of the Dzyaloshinskii-Moriya type. The parameters of the model can be flexibly tuned by change of the geometry of the lattice (antidot size and inter-dot distance), which can be routinely achieved within the state of the art fabrication techniques. The proposed system can thus serve as a solid state quantum simulator of a wide class of quantum compass models with possible applications ranging from emulation of correlated electron materials to topological quantum computation.
\begin{figure}[h]
\includegraphics[scale = 0.225]{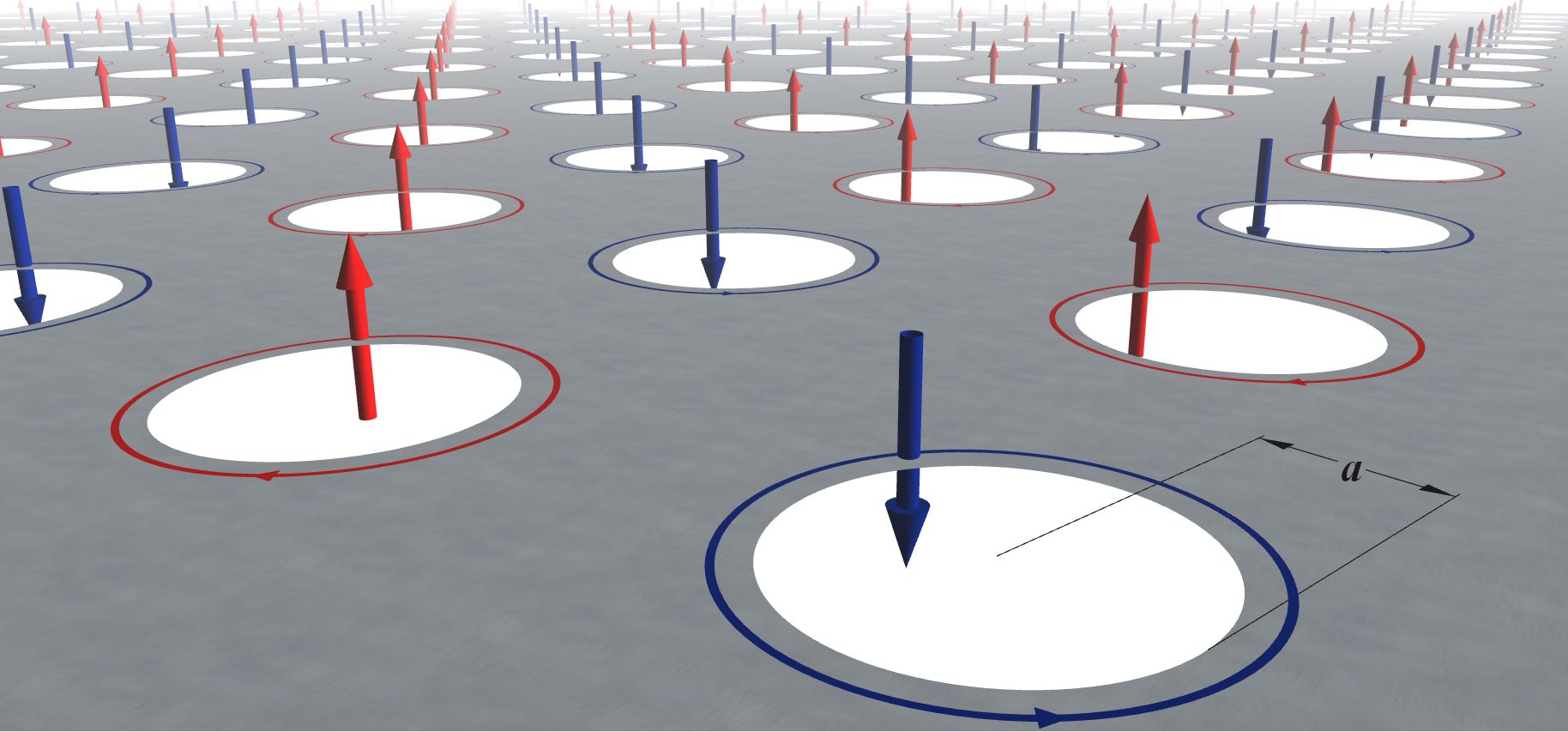}% Here is how to import EPS art
\caption{\label{fig:system}Two-dimensional array of antidots (holes) in 2DTI. Blue and red arrows show topologically-protected counterpropagating edge states with opposite spin directions.}
\end{figure}

The geometry of the  system we consider is shown schematically in Fig.~\ref{fig:system}. We take an example of CdTe/HgTe/CdTe quantum well~\cite{Bernevig_Hughes2013} since this is currently the most common of 2DTI where topologically protected edge states has been observed experimentally~\cite{Konig2008}, but other material platforms are also possible. 

The model Hamiltonian of a  CdTe/HgTe/CdTe quantum well is represented by 4 by 4 block-diagonal matrix, which consists of the blocks related to each other by time-reversal symmetry operation
\cite{Bernevig_Hughes2013}, $H=\mathrm{diag}[h(k),h^*(-k)]$, where
\begin{align}
h(k) = (C-Dk^2)+(M-Bk^2)\sigma_z+A{\sigma}\cdot\mathbf{k}
\end{align}
The parameters entering into this expression are determined by the geometry of the QW. Further, we take $A=375$ meV~nm, $B=-1.12$ eV~nm$^2$, $D=0 0$, $C=0$, $M=-10$~meV~\cite{Michetti_2011}. The parameter $M$ (Dirac mass), defined by a thickness of a QW, is of special importance, as only the case $M <$ 0 corresponds to the topologically non-trivial regime. 

The eigenvalues and eigenvectors of an individual axially symmetric antidot can be found with use of the following ansatz for a 4-spinor:
\begin{equation}
\label{wavefunction}
\psi = \frac{e^{im\theta}}{\sqrt{2\pi}}
\begin{pmatrix}
  \chi_1^m(r) e^{i\theta/2}\\
  \chi_2^m(r) e^{-i\theta/2}
\end{pmatrix},
\end{equation}
where $\theta$~is an angular coordinate, and quantum number $m = \pm1/2,\pm3/2,\pm5/2,...$ gives z-component of the total angular momentum $j_z$ commuting with the Hamiltonian. Using this substitution we get the radial part of a wavefunction in terms of the Macdonald functions $K_{m\pm 1/2}$. Applying the Dirichlet boundary condition at the antidot edge $\chi_{1,2}(r)|_{r=a}=0$ we get a secular equation defining the eigenergies and eigenfunctions. 

\begin{figure}[htb]
\includegraphics[scale = 0.245]{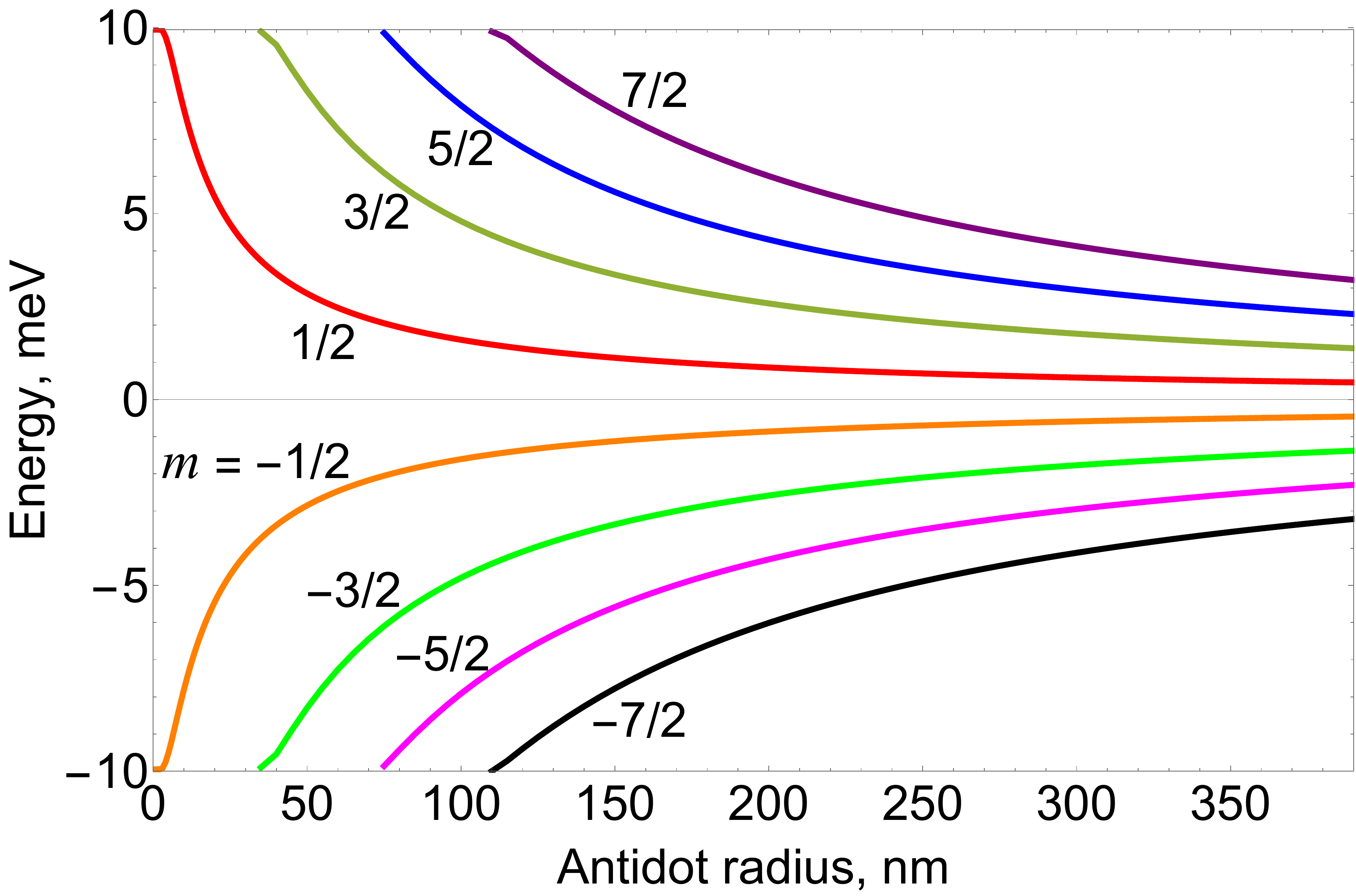}% Here is how to import EPS art
\caption{\label{energy}Energies of in-gap edge states circulating around an antidot as functions of the antidot radius $a$. $m$ is the quantum number defining z-projection of the total angular momentum of the circulating edge states.}
\end{figure}

The spectrum of a single antidot as a function of an antidot radius $a$ is shown in Fig.~\ref{energy}. As one can see, the spectrum is symmetric with respect to the gap centre (zero energy). For small radii there exist only two bound states corresponding  $m=\pm 1/2$ with energies approaching the gap edges as one decreases the radius. At larger radii the states corresponding to larger $|m|$ appear. In what follows we consider the radius $a=15$~nm, which corresponds to the case of a single pair of the bound states.

It should be stressed, that each of the eigenenergies corresponds to the Kramer's doublet, representing a mixture of purely orbital degenerate states~\cite{PhysRevLett.82.1016} and degenerate spin states. We introduce the pseudospin index $\sigma=(\uparrow~,~\downarrow)$ to label the partners of the doublet. Their wavefunctions $\psi$ and $\psi'$ are related to each other by the time reversal operator $\mathcal{T}$: $\psi'=\mathcal T \psi= i\sigma_y\psi^\ast$. 

\begin{figure}[htb]
\begin{minipage}[h]{1\linewidth}
\center{\includegraphics[scale = 0.245]{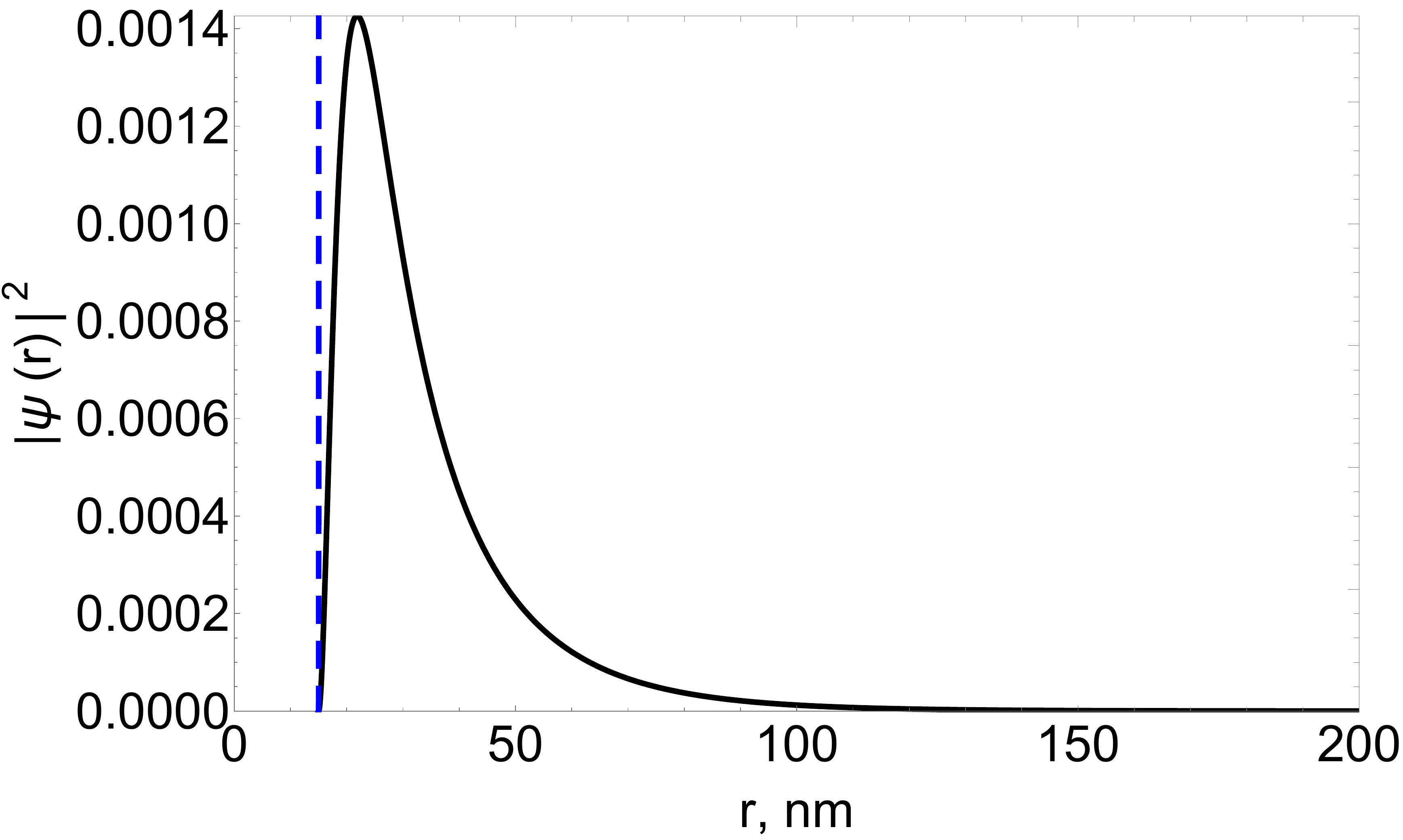}}
\end{minipage}
\put(-40,50){
  \makebox(0,0)[lb]{$(a)$}}
\vfill
\begin{minipage}[b]{1\linewidth}
\center{\includegraphics[scale = 0.43]{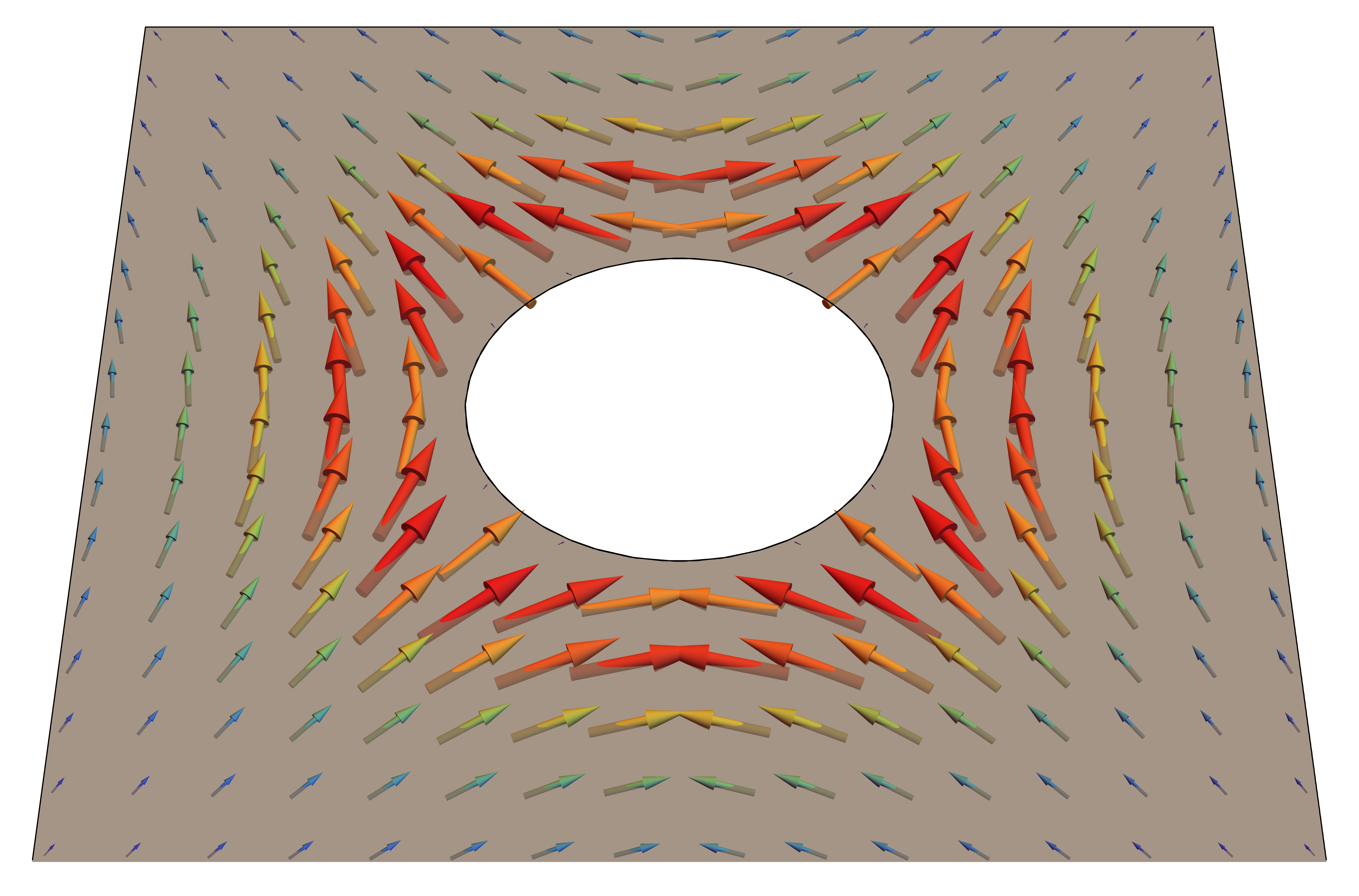}}
\end{minipage}
  \put(-15,125){
  \makebox(0,0)[lb]{$(b)$}}
\caption{(a) - The dependence of the $|\psi(r)|^2$  of the edge state of an antidot with the radius $a$ = 15~nm vs distance from the center of the antidot, $|m| = 1/2$ (b) - Spin density of the edge state near a single antidot}
\label{fig:wavefunction}
\end{figure}

In Figs.~\ref{fig:wavefunction}(a,b)  we plot the probability distribution functions and corresponding spin density profile for the $m=1/2$ eigenstate for $a=15$~nm. It can be seen that the wavefunction is localized at the scale of several $a$. Moreover the spin distribution of the state is highly non-isotropic which is a consequence of the spin-orbit coupling.

In the situation, when apertures are placed reasonably close to each other, the wavefunctions of the states localized at each of them overlap, and electrons can thus tunnel between antidots. If one neglects the doublet corresponding to higher energy (upper line in Fig. 2), the system can be described in terms of a tight binding Hamiltonian of the Hubbard type, which for a square lattice of antidots reads:
\begin{equation}
H = -\sum_{\langle i,j \rangle} (c_{i\sigma}^{+}t_{ij}^{\sigma\sigma'}c_{j\sigma'}+h.c.)+U\sum_{i} (n_{i\uparrow}-\frac{1}{2})(n_{i\downarrow}-\frac{1}{2}), \label{Hubbard}
\end{equation}
where  $c_{i\sigma}$ is the annihilation operator for the state with specific pseudospin projection localized at site $i$ of the lattice. The first term corresponds to the tunneling between the sites, while the last term describes on-site Coulomb repulsion.

\begin{figure}[!h]
\begin{minipage}[h]{1\linewidth}
\center{\includegraphics[scale = 0.1779]{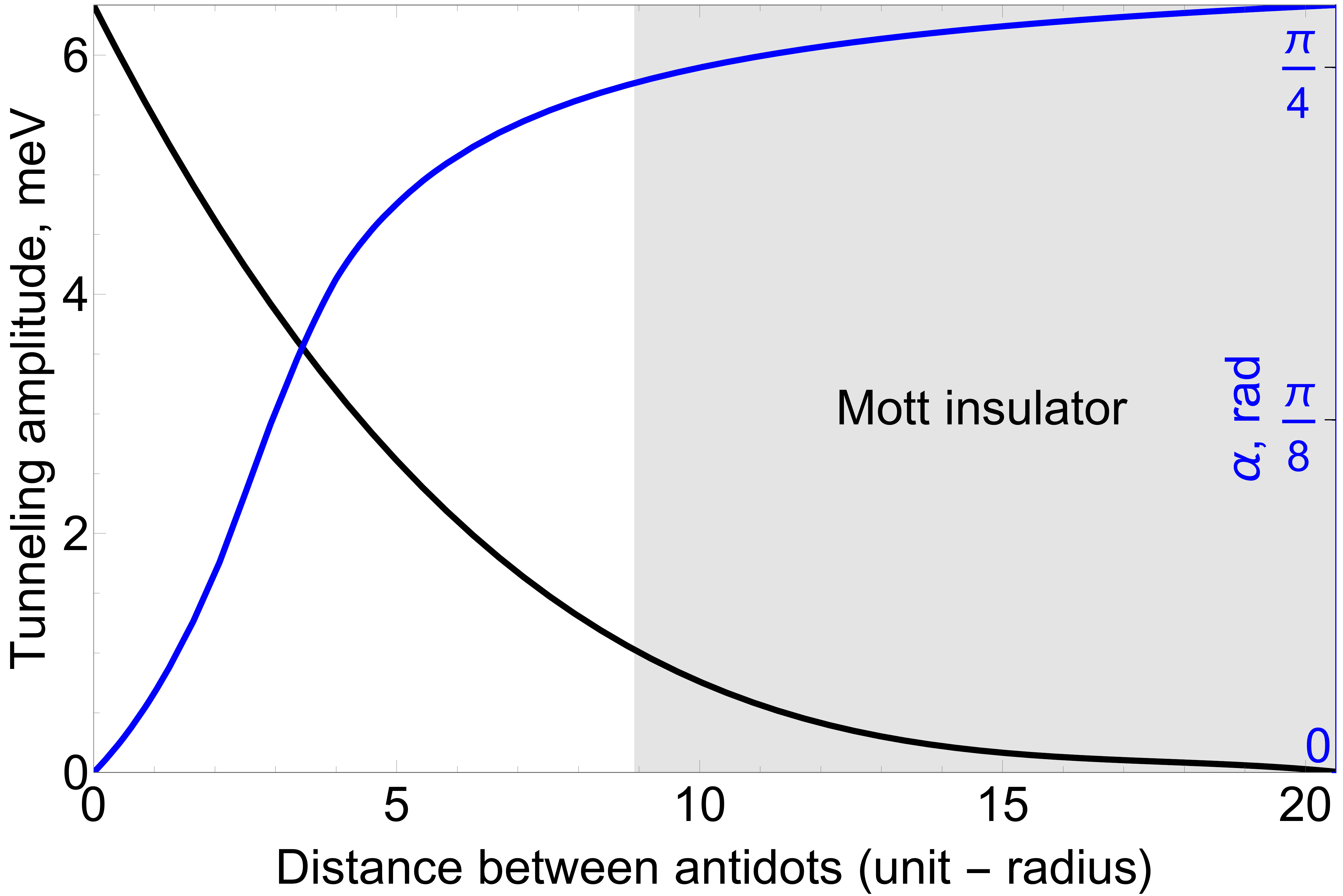}}
\end{minipage}
\put(-40.1,50){
  \makebox(0,0)[lb]{$(a)$}}
\vfill
\vspace{5 mm}
\begin{minipage}[b]{1\linewidth}
\hspace{-2mm}
\includegraphics[scale = 0.245]{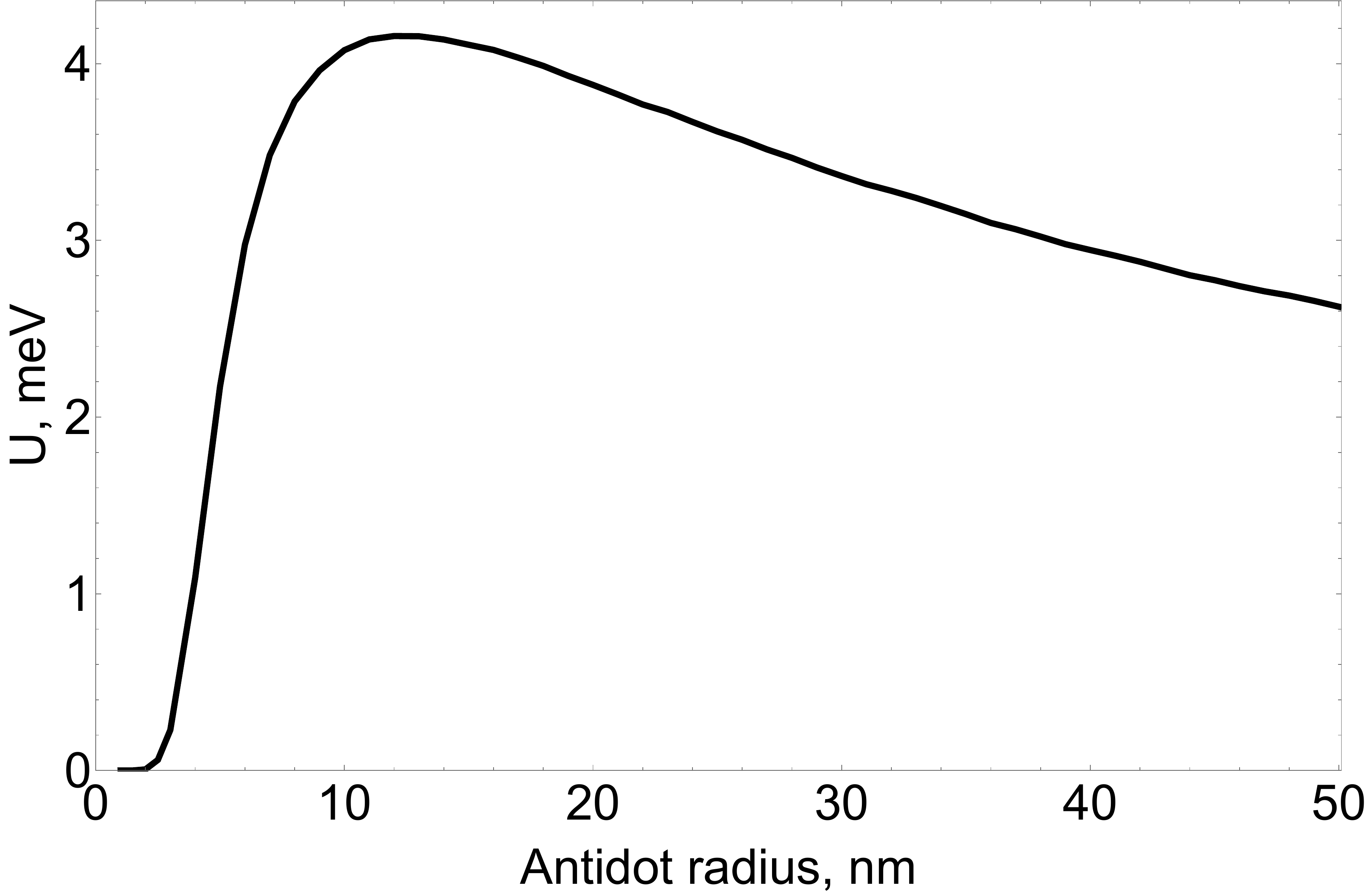}
\end{minipage}
  \put(-39.5,125){
  \makebox(0,0)[lb]{$(b)$}}
\caption{\label{tunneling}  (a) - The dependence of the tunneling amplitude (black curve) and $\alpha (\beta)$ parameter (blue curve) vs distance between centers of antidots R (in units of the dot radius $a$, taken to be 15~nm). Gray area of the plot denotes Mott insulating regime of the system at half filling. (b) - The dependence of the on-site interaction energy vs radius of the antidot.}
\label{fig:hubbard}
\end{figure}

Since we account for nearest neighbour hopping only, there exist only two inequivalent tunneling matrices $t_{i,i+\hat{x}}^{\sigma\sigma'},t_{i,i+\hat{y}}^{\sigma\sigma'}$ corresponding to hoppings along orthogonal lattice translation vectors, which read:
\begin{equation}
t_{i,i+\hat{e}}^{\sigma\sigma'}=\epsilon\int d^2\mathbf{r}\psi_{\sigma}^H(\mathbf{r})\psi_{\sigma'}(\mathbf{r}+R\hat{e}),
\end{equation}
where $R$ is the distance between the antidot centers, $\epsilon$ is the bound state energy, and $\hat{e}=[x,y]$. Substituting the expression for $\psi_{\uparrow}$ from Eq.~\eqref{wavefunction} and recalling that $\psi_{\downarrow}=\mathcal{T}\psi_{\uparrow}$, we notice that $t_{i,i+\hat{e}}^{\uparrow\uparrow}=t_{i,i+\hat{e}}^{\downarrow\downarrow}$ and $t_{i,i+\hat{e}}^{\uparrow\downarrow}=-(t_{i,i+\hat{e}}^{\downarrow\uparrow})^*$. Moreover, since $\chi_1,\chi_2$ are real functions, the diagonal elements $t^{\sigma\sigma}$ are real. Also, the absolute value of tunnelling amplitudes should  depend only on the distance between antidots and not on  the orientation of the antidot pair. These general considerations allow to parametrize the tunnelling matrix as
\begin{align}
    \hat{t}_{i,i+\hat{e}}=t\begin{pmatrix}
      \cos\alpha &  e^{i\phi_{\hat{e}}} \sin\alpha\\ -e^{i\phi_{\hat{e}}}\sin\alpha & \cos\alpha,
    \end{pmatrix}
\end{align}
where $t$ and $\alpha$ are real numbers which depend on the antidot radius and intersite distance, and phase $\phi_{\hat{e}}$ depends on the hopping direction. Numerical integration shows that $\phi_{x}=0$ and $\phi_y=\pi/2$. The dependence of the tunneling amplitude $t$ and phase $\alpha$ versus intersite distance $R$ for $a=15$~nm is shown in Fig.~\ref{fig:hubbard}(a). As expected the tunnelling amplitude decays exponentially with the inersite distance. Interestingly, the angle $\alpha$ reaches the value of $\pi/4$, which corresponds to the case of the equivalency of spin conservative and spin flip tunnelings, at some finite value of $R$.

The Coulomb interaction energy can be estimated as:
\begin{align}
U= \int\limits V(\mathbf{r}_1,\mathbf{r}_2) \rho_{\uparrow}(\mathbf{r}_1)\rho_{\downarrow}(\mathbf{r}_2)d\mathbf{r}_1d\mathbf{r}_2,
\end{align}
where $\rho_{\uparrow\downarrow}(\mathbf{r})=\psi_{\uparrow\downarrow}^{\dagger}(\mathbf{r})\psi_{\uparrow\downarrow}(\mathbf{r})$ is an electron density for the corresponding pseudospin projection, and $V$ is the interaction potential which in principle should include both static and dynamical screening. We will neglect the latter, since edge states lie in the band gap of the bulk material where there is vanishing density of the free electrons. As to the static screening, it was considered in~Ref. \cite{cui2006electrostatic} where it has been shown that the the effect of the image potential can be neglected for moderate dielectric contrasts and simple expression for the potential $V=e^2/(\varepsilon |\mathbf{r}_1-\mathbf{r}_2|)$, where $\varepsilon\approx 10$ is the static dielectric constant of HgTe can be safely used. 

In Fig.~\ref{fig:hubbard}(b) we plot the dependence of $U$ on the antidot radius $a$. Its non-monotonic behaviour can be attributed to the fact, that for small antidots the eigenfunctions are weakly localized in the radial direction, because of the approaching of the corresponding energies to the band gap edge. At the same time, for large antidot wavefunction becomes delocalized along its periphery of the radius $\approx 2a$. The interplay between these two effects defines the radius at which $U$ becomes maximal. For the considered parameters $U=U_{max}\approx 4$~meV is achieved at $a\approx 13$~nm. 

 Let us consider the situation, when the Fermi energy in the system is tuned in such a way, that one has exactly one electron per each dot (regime of half filling). It is well known, that in this situation the tunneling between the neighbouring sites can be completely blocked by the interaction. This regime corresponds to the so called spin-orbit coupled Mott insulator and is achieved when $U>4t$~\cite{jackeli2009mott}, which in our case corresponds to the distances between the antidots $R>10a$. 
 
 Our geometry resembles some previously proposed ones, where Mott insulators were realized with arrays of semiconductor quantum dots~\cite{stafford1994collective,ugajin1994mott,byrnes2008quantum}. However, there is one crucial difference, namely strong spin-orbit coupling inherent for the topological insulators. It gives an additional twist to our model, which enables to emulate much wider class of the accessible low-energy Hamiltonians.

 By employing the standard Schrieffer-Wolff transformation and excluding the states with two electrons sitting on the same sites, we can map the low-energy sector of the original Hamiltonian \ref{Hubbard} into  the following spin lattice model:
\begin{align}
\label{eff_ham}
H_{eff} = J\sum_{i} [S_{i}^{a}R^x_{ab}(2\alpha)S_{i+\hat{x}}^{b}+S_{i}^{a}R^y_{ab}(2\beta)S_{i+\hat{y}}^{b}],
\end{align}
where exchange constant $J =4t^2/U > 0$, $a, b$ = 1, 2, 3 denote components of the pseudospin operator, $\hat{x}, \hat{y}$ are basis vectors along corresponding axes and $R^x(2\alpha)$, $R^y(2\beta)$ are SO(3)
rotation matrices around $x$ and $y$ axes, respectively. The states with pseudospins $S_z=\pm 1/2$ correspond to the occupations of the partners of the Kramers doublet, other states being their linear combinations.

For the case of $\alpha=\pi/4$ which is reached for the intersite distance $\approx 10a$, we can rewrite the Hamiltonian as
\begin{align}
    &H_{eff}=J\sum_i\sum_{\hat{e}=\hat{x},\hat{y}}\left\{ S_i^eS_{i+\hat{e}}^e+  \hat{e}\cdot [\mathbf{S}_i\times \mathbf{S}_{i+\hat{e}}]\right\}, \label{eq:heff}
\end{align}
where the first term corresponds to the so-called $90^o$ spin-compass model and the second term to the conventional Dzyaloshinskii-Moriya interaction (DMI). 

The Hamiltonian in Eq.~\eqref{eq:heff} is essentially a quantum spin compass model with added DMI interaction. Pure spin compass models are usually characterized by the highly degenerate ground states, which sometimes allows for the dimensionality reduction and even the exact solution such as in the case of the Kitaev model, corresponding to the honeycomb lattice~\cite{RevModPhys.87.1}. While for the case of the square lattice, the DMI interaction is likely to lift the ground state degeneracy, it will be instructive to consider the structures with geometrical frustration, such as Lieb or Kagome lattices, and to explore the interplay between the geometrical frustration and strong spin-orbit coupling. Moreover, inclusion of the edge states characterized by larger value of $m$ opens the access to the multi-band Hubbard model and to the multi-band Mott insulators with strong spin-orbit interaction.

In the classical limit, the Hamiltonian~\eqref{eq:heff} is characterized by the spiral wave ground states~\cite{PhysRevLett.109.085303}. At the same time, quantum fluctuations can substantially modify the ground state properties of the system. Specifically,  in~\cite{jafari2008phase} the one-dimensional analog of Hamiltonian~\eqref{eq:heff} was analyzed, where the spins are aligned along $x$ axis. It has been shown  using exact diagonalization and renormalization group methods, that the spiral long range order characterized by the order parameter
\begin{align}
    C_h=\frac{1}{4N}\sum_{i=1}^N \langle S_i^z S_{i+1}^y-S_i^y S_{i+1}^z \rangle,
\end{align}
 is destroyed and only local order is preserved.

To probe the onset of different collective phases experimentally, one can resort to the measurement of the zero-bias conductance~\cite{byrnes2008quantum}, which can be performed for different filling factors, controlled by the gate voltage. To spot quantum phase transitions, one would need cryogenic temperatures $T$ satisfying the condition $T\approx 0.01 t\approx 100\mathrm{mK}$ in order to preserve the correlations from being  washed away by thermal fluctuations~\cite{hofstetter2002high,maier2005systematic}. This temperature range can be routinely achieved in the modern dilution cryostats.
\\ \indent In conclusion, we have proposed an experimentally viable quantum simulator of a spin-orbit coupled Mott insulator based on array of antidots in the two dimensional topological insulator. If antidot size is sufficiently small, the low-energy behaviour of the system can be described by the quantum spin compass model with Dzyaloshinskii-Moriya interaction. The alternative lattice geometries supporting the geometrical frustration would give access to even richer class of the available spin models, which could be of particular interest in the domains of quantum simulation and quantum computation. 
\\ \indent The authors acknowledge the support of Russian Science Foundation, grant 19-72-20120.

\bibliography{apssamp}% Produces the bibliography via BibTeX.

\providecommand{\noopsort}[1]{}\providecommand{\singleletter}[1]{#1}%
\begin{thebibliography}{36}%
\makeatletter
\providecommand \@ifxundefined [1]{%
 \@ifx{#1\undefined}
}%
\providecommand \@ifnum [1]{%
 \ifnum #1\expandafter \@firstoftwo
 \else \expandafter \@secondoftwo
 \fi
}%
\providecommand \@ifx [1]{%
 \ifx #1\expandafter \@firstoftwo
 \else \expandafter \@secondoftwo
 \fi
}%
\providecommand \natexlab [1]{#1}%
\providecommand \enquote  [1]{``#1''}%
\providecommand \bibnamefont  [1]{#1}%
\providecommand \bibfnamefont [1]{#1}%
\providecommand \citenamefont [1]{#1}%
\providecommand \href@noop [0]{\@secondoftwo}%
\providecommand \href [0]{\begingroup \@sanitize@url \@href}%
\providecommand \@href[1]{\@@startlink{#1}\@@href}%
\providecommand \@@href[1]{\endgroup#1\@@endlink}%
\providecommand \@sanitize@url [0]{\catcode `\\12\catcode `\$12\catcode
  `\&12\catcode `\#12\catcode `\^12\catcode `\_12\catcode `\%12\relax}%
\providecommand \@@startlink[1]{}%
\providecommand \@@endlink[0]{}%
\providecommand \url  [0]{\begingroup\@sanitize@url \@url }%
\providecommand \@url [1]{\endgroup\@href {#1}{\urlprefix }}%
\providecommand \urlprefix  [0]{URL }%
\providecommand \Eprint [0]{\href }%
\providecommand \doibase [0]{http://dx.doi.org/}%
\providecommand \selectlanguage [0]{\@gobble}%
\providecommand \bibinfo  [0]{\@secondoftwo}%
\providecommand \bibfield  [0]{\@secondoftwo}%
\providecommand \translation [1]{[#1]}%
\providecommand \BibitemOpen [0]{}%
\providecommand \bibitemStop [0]{}%
\providecommand \bibitemNoStop [0]{.\EOS\space}%
\providecommand \EOS [0]{\spacefactor3000\relax}%
\providecommand \BibitemShut  [1]{\csname bibitem#1\endcsname}%
\let\auto@bib@innerbib\@empty
%</preamble>
\bibitem [{\citenamefont {Lee}\ \emph {et~al.}(2006)\citenamefont {Lee},
  \citenamefont {Nagaosa},\ and\ \citenamefont {Wen}}]{Lee2006}%
  \BibitemOpen
  \bibfield  {author} {\bibinfo {author} {\bibfnamefont {Patrick~A.}\
  \bibnamefont {Lee}}, \bibinfo {author} {\bibfnamefont {Naoto}\ \bibnamefont
  {Nagaosa}}, \ and\ \bibinfo {author} {\bibfnamefont {Xiao-Gang}\ \bibnamefont
  {Wen}},\ }\bibfield  {title} {\enquote {\bibinfo {title} {Doping a Mott
  insulator: Physics of high-temperature superconductivity},}\ }\href {\doibase
  10.1103/RevModPhys.78.17} {\bibfield  {journal} {\bibinfo  {journal} {Rev.
  Mod. Phys.}\ }\textbf {\bibinfo {volume} {78}},\ \bibinfo {pages} {17--85}
  (\bibinfo {year} {2006})}\BibitemShut {NoStop}%
\bibitem [{\citenamefont {Nelson}(1979)}]{Nelson1979}%
  \BibitemOpen
  \bibfield  {author} {\bibinfo {author} {\bibfnamefont {D.~R.}\ \bibnamefont
  {Nelson}},\ }\bibfield  {title} {\enquote {\bibinfo {title} {Superfluidity
  and the two dimensional xy model},}\ }\href@noop {} {\bibfield  {journal}
  {\bibinfo  {journal} {Phys. Rep.}\ }\textbf {\bibinfo {volume} {49}},\
  \bibinfo {pages} {255} (\bibinfo {year} {1979})}\BibitemShut {NoStop}%
\bibitem [{\citenamefont {Kyriienko}\ \emph {et~al.}(2015)\citenamefont
  {Kyriienko}, \citenamefont {Wierschem}, \citenamefont {Sengupta},\ and\
  \citenamefont {Shelykh}}]{Kyriienko2015}%
  \BibitemOpen
  \bibfield  {author} {\bibinfo {author} {\bibfnamefont {O.}~\bibnamefont
  {Kyriienko}}, \bibinfo {author} {\bibfnamefont {K.}~\bibnamefont
  {Wierschem}}, \bibinfo {author} {\bibfnamefont {P.}~\bibnamefont {Sengupta}},
  \ and\ \bibinfo {author} {\bibfnamefont {I.~A.}\ \bibnamefont {Shelykh}},\
  }\bibfield  {title} {\enquote {\bibinfo {title} {Quantum hall bilayer as
  pseudospin magnet},}\ }\href@noop {} {\bibfield  {journal} {\bibinfo
  {journal} {Europhys. Lett.}\ }\textbf {\bibinfo {volume} {109}},\ \bibinfo
  {pages} {57003} (\bibinfo {year} {2015})}\BibitemShut {NoStop}%
\bibitem [{\citenamefont {Kalinin}\ \emph {et~al.}(2020)\citenamefont
  {Kalinin}, \citenamefont {Amo}, \citenamefont {Bloch},\ and\ \citenamefont
  {Berloff}}]{Kalinin2020}%
  \BibitemOpen
  \bibfield  {author} {\bibinfo {author} {\bibfnamefont {K.~P.}\ \bibnamefont
  {Kalinin}}, \bibinfo {author} {\bibfnamefont {A.}~\bibnamefont {Amo}},
  \bibinfo {author} {\bibfnamefont {J.}~\bibnamefont {Bloch}}, \ and\ \bibinfo
  {author} {\bibfnamefont {N.~G.}\ \bibnamefont {Berloff}},\ }\bibfield
  {title} {\enquote {\bibinfo {title} {Quantum hall bilayer as pseudospin
  magnet},}\ }\href@noop {} {\bibfield  {journal} {\bibinfo  {journal} {Nat.
  Photon.}\ }\textbf {\bibinfo {volume} {9}},\ \bibinfo {pages} {4127}
  (\bibinfo {year} {2020})}\BibitemShut {NoStop}%
\bibitem [{\citenamefont {Nussinov}\ and\ \citenamefont {van~den
  Brink}(2015)}]{RevModPhys.87.1}%
  \BibitemOpen
  \bibfield  {author} {\bibinfo {author} {\bibfnamefont {Zohar}\ \bibnamefont
  {Nussinov}}\ and\ \bibinfo {author} {\bibfnamefont {Jeroen}\ \bibnamefont
  {van~den Brink}},\ }\bibfield  {title} {\enquote {\bibinfo {title} {Compass
  models: Theory and physical motivations},}\ }\href {\doibase
  10.1103/RevModPhys.87.1} {\bibfield  {journal} {\bibinfo  {journal} {Rev.
  Mod. Phys.}\ }\textbf {\bibinfo {volume} {87}},\ \bibinfo {pages} {1--59}
  (\bibinfo {year} {2015})}\BibitemShut {NoStop}%
\bibitem [{\citenamefont {Kugel}\ and\ \citenamefont
  {Khomski{\u{\i}}}(1982)}]{kugel1982jahn}%
  \BibitemOpen
  \bibfield  {author} {\bibinfo {author} {\bibfnamefont {Kliment~I}\
  \bibnamefont {Kugel}}\ and\ \bibinfo {author} {\bibfnamefont
  {DI}~\bibnamefont {Khomski{\u{\i}}}},\ }\bibfield  {title} {\enquote
  {\bibinfo {title} {The jahn-teller effect and magnetism: transition metal
  compounds},}\ }\href@noop {} {\bibfield  {journal} {\bibinfo  {journal}
  {Soviet Physics Uspekhi}\ }\textbf {\bibinfo {volume} {25}},\ \bibinfo
  {pages} {231} (\bibinfo {year} {1982})}\BibitemShut {NoStop}%
\bibitem [{\citenamefont {Kuroki}\ \emph {et~al.}(2008)\citenamefont {Kuroki},
  \citenamefont {Onari}, \citenamefont {Arita}, \citenamefont {Usui},
  \citenamefont {Tanaka}, \citenamefont {Kontani},\ and\ \citenamefont
  {Aoki}}]{PhysRevLett.101.087004}%
  \BibitemOpen
  \bibfield  {author} {\bibinfo {author} {\bibfnamefont {Kazuhiko}\
  \bibnamefont {Kuroki}}, \bibinfo {author} {\bibfnamefont {Seiichiro}\
  \bibnamefont {Onari}}, \bibinfo {author} {\bibfnamefont {Ryotaro}\
  \bibnamefont {Arita}}, \bibinfo {author} {\bibfnamefont {Hidetomo}\
  \bibnamefont {Usui}}, \bibinfo {author} {\bibfnamefont {Yukio}\ \bibnamefont
  {Tanaka}}, \bibinfo {author} {\bibfnamefont {Hiroshi}\ \bibnamefont
  {Kontani}}, \ and\ \bibinfo {author} {\bibfnamefont {Hideo}\ \bibnamefont
  {Aoki}},\ }\bibfield  {title} {\enquote {\bibinfo {title} {Unconventional
  pairing originating from the disconnected Fermi surfaces of superconducting
  ${\mathrm{lafeaso}}_{1\ensuremath{-}x}{\mathrm{f}}_{x}$},}\ }\href {\doibase
  10.1103/PhysRevLett.101.087004} {\bibfield  {journal} {\bibinfo  {journal}
  {Phys. Rev. Lett.}\ }\textbf {\bibinfo {volume} {101}},\ \bibinfo {pages}
  {087004} (\bibinfo {year} {2008})}\BibitemShut {NoStop}%
\bibitem [{\citenamefont {Kr\"uger}\ \emph {et~al.}(2009)\citenamefont
  {Kr\"uger}, \citenamefont {Kumar}, \citenamefont {Zaanen},\ and\
  \citenamefont {van~den Brink}}]{PhysRevB.79.054504}%
  \BibitemOpen
  \bibfield  {author} {\bibinfo {author} {\bibfnamefont {Frank}\ \bibnamefont
  {Kr\"uger}}, \bibinfo {author} {\bibfnamefont {Sanjeev}\ \bibnamefont
  {Kumar}}, \bibinfo {author} {\bibfnamefont {Jan}\ \bibnamefont {Zaanen}}, \
  and\ \bibinfo {author} {\bibfnamefont {Jeroen}\ \bibnamefont {van~den
  Brink}},\ }\bibfield  {title} {\enquote {\bibinfo {title} {Spin-orbital
  frustrations and anomalous metallic state in iron-pnictide
  superconductors},}\ }\href {\doibase 10.1103/PhysRevB.79.054504} {\bibfield
  {journal} {\bibinfo  {journal} {Phys. Rev. B}\ }\textbf {\bibinfo {volume}
  {79}},\ \bibinfo {pages} {054504} (\bibinfo {year} {2009})}\BibitemShut
  {NoStop}%
\bibitem [{\citenamefont {Trousselet}\ \emph {et~al.}(2012)\citenamefont
  {Trousselet}, \citenamefont {Ole\ifmmode~\acute{s}\else \'{s}\fi{}},\ and\
  \citenamefont {Horsch}}]{PhysRevB.86.134412}%
  \BibitemOpen
  \bibfield  {author} {\bibinfo {author} {\bibfnamefont {Fabien}\ \bibnamefont
  {Trousselet}}, \bibinfo {author} {\bibfnamefont {Andrzej~M.}\ \bibnamefont
  {Ole\ifmmode~\acute{s}\else \'{s}\fi{}}}, \ and\ \bibinfo {author}
  {\bibfnamefont {Peter}\ \bibnamefont {Horsch}},\ }\bibfield  {title}
  {\enquote {\bibinfo {title} {Magnetic properties of nanoscale
  compass-heisenberg planar clusters},}\ }\href {\doibase
  10.1103/PhysRevB.86.134412} {\bibfield  {journal} {\bibinfo  {journal} {Phys.
  Rev. B}\ }\textbf {\bibinfo {volume} {86}},\ \bibinfo {pages} {134412}
  (\bibinfo {year} {2012})}\BibitemShut {NoStop}%
\bibitem [{\citenamefont {Tokura}\ and\ \citenamefont
  {Tomioka}(1999)}]{tokura1999colossal}%
  \BibitemOpen
  \bibfield  {author} {\bibinfo {author} {\bibfnamefont {Yoshinori}\
  \bibnamefont {Tokura}}\ and\ \bibinfo {author} {\bibfnamefont {Yasuhide}\
  \bibnamefont {Tomioka}},\ }\bibfield  {title} {\enquote {\bibinfo {title}
  {Colossal magnetoresistive manganites},}\ }\href@noop {} {\bibfield
  {journal} {\bibinfo  {journal} {Journal of magnetism and magnetic materials}\
  }\textbf {\bibinfo {volume} {200}},\ \bibinfo {pages} {1--23} (\bibinfo
  {year} {1999})}\BibitemShut {NoStop}%
\bibitem [{\citenamefont {Baek}\ \emph {et~al.}(2017)\citenamefont {Baek},
  \citenamefont {Do}, \citenamefont {Choi}, \citenamefont {Kwon}, \citenamefont
  {Wolter}, \citenamefont {Nishimoto}, \citenamefont {Van Den~Brink},\ and\
  \citenamefont {B{\"u}chner}}]{baek2017evidence}%
  \BibitemOpen
  \bibfield  {author} {\bibinfo {author} {\bibfnamefont {S-H}\ \bibnamefont
  {Baek}}, \bibinfo {author} {\bibfnamefont {S-H}\ \bibnamefont {Do}}, \bibinfo
  {author} {\bibfnamefont {K-Y}\ \bibnamefont {Choi}}, \bibinfo {author}
  {\bibfnamefont {Yong~Seung}\ \bibnamefont {Kwon}}, \bibinfo {author}
  {\bibfnamefont {AUB}\ \bibnamefont {Wolter}}, \bibinfo {author}
  {\bibfnamefont {S}~\bibnamefont {Nishimoto}}, \bibinfo {author}
  {\bibfnamefont {Jeroen}\ \bibnamefont {Van Den~Brink}}, \ and\ \bibinfo
  {author} {\bibfnamefont {B}~\bibnamefont {B{\"u}chner}},\ }\bibfield  {title}
  {\enquote {\bibinfo {title} {Evidence for a field-induced quantum spin liquid
  in $\alpha$-rucl 3},}\ }\href@noop {} {\bibfield  {journal} {\bibinfo
  {journal} {Physical Review Letters}\ }\textbf {\bibinfo {volume} {119}},\
  \bibinfo {pages} {037201} (\bibinfo {year} {2017})}\BibitemShut {NoStop}%
\bibitem [{\citenamefont {Kitaev}(2003)}]{kitaev2003fault}%
  \BibitemOpen
  \bibfield  {author} {\bibinfo {author} {\bibfnamefont {A~Yu}\ \bibnamefont
  {Kitaev}},\ }\bibfield  {title} {\enquote {\bibinfo {title} {Fault-tolerant
  quantum computation by anyons},}\ }\href@noop {} {\bibfield  {journal}
  {\bibinfo  {journal} {Annals of Physics}\ }\textbf {\bibinfo {volume}
  {303}},\ \bibinfo {pages} {2--30} (\bibinfo {year} {2003})}\BibitemShut
  {NoStop}%
\bibitem [{\citenamefont {Cubitt}\ \emph {et~al.}(2018)\citenamefont {Cubitt},
  \citenamefont {Montanaro},\ and\ \citenamefont
  {Piddock}}]{cubitt2018universal}%
  \BibitemOpen
  \bibfield  {author} {\bibinfo {author} {\bibfnamefont {Toby~S}\ \bibnamefont
  {Cubitt}}, \bibinfo {author} {\bibfnamefont {Ashley}\ \bibnamefont
  {Montanaro}}, \ and\ \bibinfo {author} {\bibfnamefont {Stephen}\ \bibnamefont
  {Piddock}},\ }\bibfield  {title} {\enquote {\bibinfo {title} {Universal
  quantum hamiltonians},}\ }\href@noop {} {\bibfield  {journal} {\bibinfo
  {journal} {Proceedings of the National Academy of Sciences}\ }\textbf
  {\bibinfo {volume} {115}},\ \bibinfo {pages} {9497--9502} (\bibinfo {year}
  {2018})}\BibitemShut {NoStop}%
\bibitem [{\citenamefont {Wu}(2008)}]{wu2008orbital}%
  \BibitemOpen
  \bibfield  {author} {\bibinfo {author} {\bibfnamefont {Congjun}\ \bibnamefont
  {Wu}},\ }\bibfield  {title} {\enquote {\bibinfo {title} {Orbital ordering and
  frustration of p-band Mott insulators},}\ }\href@noop {} {\bibfield
  {journal} {\bibinfo  {journal} {Physical Review Letters}\ }\textbf {\bibinfo
  {volume} {100}},\ \bibinfo {pages} {200406} (\bibinfo {year}
  {2008})}\BibitemShut {NoStop}%
\bibitem [{\citenamefont {Radi{\'c}}\ \emph {et~al.}(2012)\citenamefont
  {Radi{\'c}}, \citenamefont {Di~Ciolo}, \citenamefont {Sun},\ and\
  \citenamefont {Galitski}}]{radic2012exotic}%
  \BibitemOpen
  \bibfield  {author} {\bibinfo {author} {\bibfnamefont {J}~\bibnamefont
  {Radi{\'c}}}, \bibinfo {author} {\bibfnamefont {Andrea}\ \bibnamefont
  {Di~Ciolo}}, \bibinfo {author} {\bibfnamefont {Kai}\ \bibnamefont {Sun}}, \
  and\ \bibinfo {author} {\bibfnamefont {Victor}\ \bibnamefont {Galitski}},\
  }\bibfield  {title} {\enquote {\bibinfo {title} {Exotic quantum spin models
  in spin-orbit-coupled Mott insulators},}\ }\href@noop {} {\bibfield
  {journal} {\bibinfo  {journal} {Physical Review Letters}\ }\textbf {\bibinfo
  {volume} {109}},\ \bibinfo {pages} {085303} (\bibinfo {year}
  {2012})}\BibitemShut {NoStop}%
\bibitem [{\citenamefont {Hasan}\ and\ \citenamefont
  {Kane}(2010)}]{RevModPhys.82.3045}%
  \BibitemOpen
  \bibfield  {author} {\bibinfo {author} {\bibfnamefont {M.~Z.}\ \bibnamefont
  {Hasan}}\ and\ \bibinfo {author} {\bibfnamefont {C.~L.}\ \bibnamefont
  {Kane}},\ }\bibfield  {title} {\enquote {\bibinfo {title} {Colloquium:
  Topological insulators},}\ }\href {\doibase 10.1103/RevModPhys.82.3045}
  {\bibfield  {journal} {\bibinfo  {journal} {Rev. Mod. Phys.}\ }\textbf
  {\bibinfo {volume} {82}},\ \bibinfo {pages} {3045--3067} (\bibinfo {year}
  {2010})}\BibitemShut {NoStop}%
\bibitem [{\citenamefont {Kane}(2013)}]{KANE20133}%
  \BibitemOpen
  \bibfield  {author} {\bibinfo {author} {\bibfnamefont {C.L.}\ \bibnamefont
  {Kane}},\ }\bibfield  {title} {\enquote {\bibinfo {title} {Chapter 1 -
  topological band theory and the Z2 invariant},}\ }in\ \href {\doibase
  https://doi.org/10.1016/B978-0-444-63314-9.00001-9} {\emph {\bibinfo
  {booktitle} {Topological Insulators}}},\ \bibinfo {series} {Contemporary
  Concepts of Condensed Matter Science}, Vol.~\bibinfo {volume} {6},\ \bibinfo
  {editor} {edited by\ \bibinfo {editor} {\bibfnamefont {Marcel}\ \bibnamefont
  {Franz}}\ and\ \bibinfo {editor} {\bibfnamefont {Laurens}\ \bibnamefont
  {Molenkamp}}}\ (\bibinfo  {publisher} {Elsevier},\ \bibinfo {year} {2013})\
  pp.\ \bibinfo {pages} {3 -- 34}\BibitemShut {NoStop}%
\bibitem [{\citenamefont {Bernevig}\ and\ \citenamefont
  {Hughes}(2013)}]{Bernevig_Hughes2013}%
  \BibitemOpen
  \bibfield  {author} {\bibinfo {author} {\bibfnamefont {B.~Andrei.}\
  \bibnamefont {Bernevig}}\ and\ \bibinfo {author} {\bibfnamefont {Taylor~L.}\
  \bibnamefont {Hughes}},\ }\href@noop {} {\emph {\bibinfo {title} {Topological
  Insulators and Topological Superconductors}}}\ (\bibinfo  {publisher}
  {Princeton Univ. Press},\ \bibinfo {year} {2013})\BibitemShut {NoStop}%
\bibitem [{\citenamefont {Barik}\ \emph {et~al.}(2020)\citenamefont {Barik},
  \citenamefont {Karasahin}, \citenamefont {Mittal}, \citenamefont {Waks},\
  and\ \citenamefont {Hafezi}}]{PhysRevB.101.205303}%
  \BibitemOpen
  \bibfield  {author} {\bibinfo {author} {\bibfnamefont {Sabyasachi}\
  \bibnamefont {Barik}}, \bibinfo {author} {\bibfnamefont {Aziz}\ \bibnamefont
  {Karasahin}}, \bibinfo {author} {\bibfnamefont {Sunil}\ \bibnamefont
  {Mittal}}, \bibinfo {author} {\bibfnamefont {Edo}\ \bibnamefont {Waks}}, \
  and\ \bibinfo {author} {\bibfnamefont {Mohammad}\ \bibnamefont {Hafezi}},\
  }\bibfield  {title} {\enquote {\bibinfo {title} {Chiral quantum optics using
  a topological resonator},}\ }\href {\doibase 10.1103/PhysRevB.101.205303}
  {\bibfield  {journal} {\bibinfo  {journal} {Phys. Rev. B}\ }\textbf {\bibinfo
  {volume} {101}},\ \bibinfo {pages} {205303} (\bibinfo {year}
  {2020})}\BibitemShut {NoStop}%
\bibitem [{\citenamefont {Jalali~Mehrabad}\ \emph {et~al.}(2020)\citenamefont
  {Jalali~Mehrabad}, \citenamefont {Foster}, \citenamefont {Dost},
  \citenamefont {Clarke}, \citenamefont {Patil}, \citenamefont {Farrer},
  \citenamefont {Heffernan}, \citenamefont {Skolnick},\ and\ \citenamefont
  {Wilson}}]{jalali2020semiconductor}%
  \BibitemOpen
  \bibfield  {author} {\bibinfo {author} {\bibfnamefont {M}~\bibnamefont
  {Jalali~Mehrabad}}, \bibinfo {author} {\bibfnamefont {AP}~\bibnamefont
  {Foster}}, \bibinfo {author} {\bibfnamefont {R}~\bibnamefont {Dost}},
  \bibinfo {author} {\bibfnamefont {E}~\bibnamefont {Clarke}}, \bibinfo
  {author} {\bibfnamefont {PK}~\bibnamefont {Patil}}, \bibinfo {author}
  {\bibfnamefont {I}~\bibnamefont {Farrer}}, \bibinfo {author} {\bibfnamefont
  {J}~\bibnamefont {Heffernan}}, \bibinfo {author} {\bibfnamefont
  {MS}~\bibnamefont {Skolnick}}, \ and\ \bibinfo {author} {\bibfnamefont
  {LR}~\bibnamefont {Wilson}},\ }\bibfield  {title} {\enquote {\bibinfo {title}
  {A semiconductor topological photonic ring resonator},}\ }\href@noop {}
  {\bibfield  {journal} {\bibinfo  {journal} {Applied Physics Letters}\
  }\textbf {\bibinfo {volume} {116}},\ \bibinfo {pages} {061102} (\bibinfo
  {year} {2020})}\BibitemShut {NoStop}%
\bibitem [{\citenamefont {Mehrabad}\ \emph {et~al.}(2020)\citenamefont
  {Mehrabad}, \citenamefont {Foster}, \citenamefont {Dost}, \citenamefont
  {Clarke}, \citenamefont {Patil}, \citenamefont {Fox}, \citenamefont
  {Skolnick},\ and\ \citenamefont {Wilson}}]{mehrabad2020chiral}%
  \BibitemOpen
  \bibfield  {author} {\bibinfo {author} {\bibfnamefont {Mahmoud~Jalali}\
  \bibnamefont {Mehrabad}}, \bibinfo {author} {\bibfnamefont {Andrew~P}\
  \bibnamefont {Foster}}, \bibinfo {author} {\bibfnamefont {Ren{\'e}}\
  \bibnamefont {Dost}}, \bibinfo {author} {\bibfnamefont {Edmund}\ \bibnamefont
  {Clarke}}, \bibinfo {author} {\bibfnamefont {Pallavi~K}\ \bibnamefont
  {Patil}}, \bibinfo {author} {\bibfnamefont {A~Mark}\ \bibnamefont {Fox}},
  \bibinfo {author} {\bibfnamefont {Maurice~S}\ \bibnamefont {Skolnick}}, \
  and\ \bibinfo {author} {\bibfnamefont {Luke~R}\ \bibnamefont {Wilson}},\
  }\bibfield  {title} {\enquote {\bibinfo {title} {Chiral topological photonics
  with an embedded quantum emitter},}\ }\href@noop {} {\bibfield  {journal}
  {\bibinfo  {journal} {Optica}\ }\textbf {\bibinfo {volume} {7}},\ \bibinfo
  {pages} {1690--1696} (\bibinfo {year} {2020})}\BibitemShut {NoStop}%
\bibitem [{\citenamefont {Michetti}\ and\ \citenamefont
  {Recher}(2011)}]{Michetti_2011}%
  \BibitemOpen
  \bibfield  {author} {\bibinfo {author} {\bibfnamefont {Paolo}\ \bibnamefont
  {Michetti}}\ and\ \bibinfo {author} {\bibfnamefont {Patrik}\ \bibnamefont
  {Recher}},\ }\bibfield  {title} {\enquote {\bibinfo {title} {Bound states and
  persistent currents in topological insulator rings},}\ }\href {\doibase
  10.1103/physrevb.83.125420} {\bibfield  {journal} {\bibinfo  {journal}
  {Physical Review B}\ }\textbf {\bibinfo {volume} {83}} (\bibinfo {year}
  {2011}),\ 10.1103/physrevb.83.125420}\BibitemShut {NoStop}%
\bibitem [{\citenamefont {Shan}\ \emph {et~al.}(2011)\citenamefont {Shan},
  \citenamefont {Lu}, \citenamefont {Lu},\ and\ \citenamefont
  {Shen}}]{PhysRevB.84.035307}%
  \BibitemOpen
  \bibfield  {author} {\bibinfo {author} {\bibfnamefont {Wen-Yu}\ \bibnamefont
  {Shan}}, \bibinfo {author} {\bibfnamefont {Jie}\ \bibnamefont {Lu}}, \bibinfo
  {author} {\bibfnamefont {Hai-Zhou}\ \bibnamefont {Lu}}, \ and\ \bibinfo
  {author} {\bibfnamefont {Shun-Qing}\ \bibnamefont {Shen}},\ }\bibfield
  {title} {\enquote {\bibinfo {title} {Vacancy-induced bound states in
  topological insulators},}\ }\href {\doibase 10.1103/PhysRevB.84.035307}
  {\bibfield  {journal} {\bibinfo  {journal} {Phys. Rev. B}\ }\textbf {\bibinfo
  {volume} {84}},\ \bibinfo {pages} {035307} (\bibinfo {year}
  {2011})}\BibitemShut {NoStop}%
\bibitem [{\citenamefont {L{\"u}}\ \emph {et~al.}(2013)\citenamefont {L{\"u}},
  \citenamefont {Lu}, \citenamefont {Shen},\ and\ \citenamefont
  {Ng}}]{lu2013quantum}%
  \BibitemOpen
  \bibfield  {author} {\bibinfo {author} {\bibfnamefont {Hai-Feng}\
  \bibnamefont {L{\"u}}}, \bibinfo {author} {\bibfnamefont {Hai-Zhou}\
  \bibnamefont {Lu}}, \bibinfo {author} {\bibfnamefont {Shun-Qing}\
  \bibnamefont {Shen}}, \ and\ \bibinfo {author} {\bibfnamefont {Tai-Kai}\
  \bibnamefont {Ng}},\ }\bibfield  {title} {\enquote {\bibinfo {title} {Quantum
  impurity in the bulk of a topological insulator},}\ }\href@noop {} {\bibfield
   {journal} {\bibinfo  {journal} {Physical Review B}\ }\textbf {\bibinfo
  {volume} {87}},\ \bibinfo {pages} {195122} (\bibinfo {year}
  {2013})}\BibitemShut {NoStop}%
\bibitem [{\citenamefont {Xin}\ and\ \citenamefont
  {Zhou}(2015)}]{xin2015kondo}%
  \BibitemOpen
  \bibfield  {author} {\bibinfo {author} {\bibfnamefont {Xianhao}\ \bibnamefont
  {Xin}}\ and\ \bibinfo {author} {\bibfnamefont {Di}~\bibnamefont {Zhou}},\
  }\bibfield  {title} {\enquote {\bibinfo {title} {Kondo effect in a
  topological insulator quantum dot},}\ }\href@noop {} {\bibfield  {journal}
  {\bibinfo  {journal} {Physical Review B}\ }\textbf {\bibinfo {volume} {91}},\
  \bibinfo {pages} {165120} (\bibinfo {year} {2015})}\BibitemShut {NoStop}%
\bibitem [{\citenamefont {König}\ \emph {et~al.}(2008)\citenamefont {König},
  \citenamefont {Buhmann}, \citenamefont {W.~Molenkamp}, \citenamefont
  {Hughes}, \citenamefont {Liu}, \citenamefont {Qi},\ and\ \citenamefont
  {Zhang}}]{Konig2008}%
  \BibitemOpen
  \bibfield  {author} {\bibinfo {author} {\bibfnamefont {Markus}\ \bibnamefont
  {König}}, \bibinfo {author} {\bibfnamefont {Hartmut}\ \bibnamefont
  {Buhmann}}, \bibinfo {author} {\bibfnamefont {Laurens}\ \bibnamefont
  {W.~Molenkamp}}, \bibinfo {author} {\bibfnamefont {Taylor}\ \bibnamefont
  {Hughes}}, \bibinfo {author} {\bibfnamefont {Chao-Xing}\ \bibnamefont {Liu}},
  \bibinfo {author} {\bibfnamefont {Xiao-Liang}\ \bibnamefont {Qi}}, \ and\
  \bibinfo {author} {\bibfnamefont {Shou-Cheng}\ \bibnamefont {Zhang}},\
  }\bibfield  {title} {\enquote {\bibinfo {title} {The quantum spin hall
  effect: Theory and experiment},}\ }\href {\doibase 10.1143/JPSJ.77.031007}
  {\bibfield  {journal} {\bibinfo  {journal} {Journal of the Physical Society
  of Japan}\ }\textbf {\bibinfo {volume} {77}},\ \bibinfo {pages} {031007}
  (\bibinfo {year} {2008})},\ \Eprint
  {http://arxiv.org/abs/https://doi.org/10.1143/JPSJ.77.031007}
  {https://doi.org/10.1143/JPSJ.77.031007} \BibitemShut {NoStop}%
\bibitem [{\citenamefont {van~den Brink}\ and\ \citenamefont
  {Khomskii}(1999)}]{PhysRevLett.82.1016}%
  \BibitemOpen
  \bibfield  {author} {\bibinfo {author} {\bibfnamefont {Jeroen}\ \bibnamefont
  {van~den Brink}}\ and\ \bibinfo {author} {\bibfnamefont {Daniel}\
  \bibnamefont {Khomskii}},\ }\bibfield  {title} {\enquote {\bibinfo {title}
  {Double exchange via degenerate orbitals},}\ }\href {\doibase
  10.1103/PhysRevLett.82.1016} {\bibfield  {journal} {\bibinfo  {journal}
  {Phys. Rev. Lett.}\ }\textbf {\bibinfo {volume} {82}},\ \bibinfo {pages}
  {1016--1019} (\bibinfo {year} {1999})}\BibitemShut {NoStop}%
\bibitem [{\citenamefont {Cui}(2006)}]{cui2006electrostatic}%
  \BibitemOpen
  \bibfield  {author} {\bibinfo {author} {\bibfnamefont {ST}~\bibnamefont
  {Cui}},\ }\bibfield  {title} {\enquote {\bibinfo {title} {Electrostatic
  potential in cylindrical dielectric media using the image charge method},}\
  }\href@noop {} {\bibfield  {journal} {\bibinfo  {journal} {Molecular
  physics}\ }\textbf {\bibinfo {volume} {104}},\ \bibinfo {pages} {2993--3001}
  (\bibinfo {year} {2006})}\BibitemShut {NoStop}%
\bibitem [{\citenamefont {Jackeli}\ and\ \citenamefont
  {Khaliullin}(2009)}]{jackeli2009mott}%
  \BibitemOpen
  \bibfield  {author} {\bibinfo {author} {\bibfnamefont {George}\ \bibnamefont
  {Jackeli}}\ and\ \bibinfo {author} {\bibfnamefont {Giniyat}\ \bibnamefont
  {Khaliullin}},\ }\bibfield  {title} {\enquote {\bibinfo {title} {Mott
  insulators in the strong spin-orbit coupling limit: from heisenberg to a
  quantum compass and Kitaev models},}\ }\href@noop {} {\bibfield  {journal}
  {\bibinfo  {journal} {Physical Review Letters}\ }\textbf {\bibinfo {volume}
  {102}},\ \bibinfo {pages} {017205} (\bibinfo {year} {2009})}\BibitemShut
  {NoStop}%
\bibitem [{\citenamefont {Stafford}\ and\ \citenamefont
  {Sarma}(1994)}]{stafford1994collective}%
  \BibitemOpen
  \bibfield  {author} {\bibinfo {author} {\bibfnamefont {CA}~\bibnamefont
  {Stafford}}\ and\ \bibinfo {author} {\bibfnamefont {S~Das}\ \bibnamefont
  {Sarma}},\ }\bibfield  {title} {\enquote {\bibinfo {title} {Collective
  coulomb blockade in an array of quantum dots: A Mott-Hubbard approach},}\
  }\href@noop {} {\bibfield  {journal} {\bibinfo  {journal} {Physical review
  letters}\ }\textbf {\bibinfo {volume} {72}},\ \bibinfo {pages} {3590}
  (\bibinfo {year} {1994})}\BibitemShut {NoStop}%
\bibitem [{\citenamefont {Ugajin}(1994)}]{ugajin1994mott}%
  \BibitemOpen
  \bibfield  {author} {\bibinfo {author} {\bibfnamefont {Ryuichi}\ \bibnamefont
  {Ugajin}},\ }\bibfield  {title} {\enquote {\bibinfo {title} {Mott
  metal-insulator transition driven by an external electric field in coupled
  quantum dot arrays and its application to field effect devices},}\
  }\href@noop {} {\bibfield  {journal} {\bibinfo  {journal} {Journal of applied
  physics}\ }\textbf {\bibinfo {volume} {76}},\ \bibinfo {pages} {2833--2836}
  (\bibinfo {year} {1994})}\BibitemShut {NoStop}%
\bibitem [{\citenamefont {Byrnes}\ \emph {et~al.}(2008)\citenamefont {Byrnes},
  \citenamefont {Kim}, \citenamefont {Kusudo},\ and\ \citenamefont
  {Yamamoto}}]{byrnes2008quantum}%
  \BibitemOpen
  \bibfield  {author} {\bibinfo {author} {\bibfnamefont {Tim}\ \bibnamefont
  {Byrnes}}, \bibinfo {author} {\bibfnamefont {Na~Young}\ \bibnamefont {Kim}},
  \bibinfo {author} {\bibfnamefont {Kenichiro}\ \bibnamefont {Kusudo}}, \ and\
  \bibinfo {author} {\bibfnamefont {Yoshihisa}\ \bibnamefont {Yamamoto}},\
  }\bibfield  {title} {\enquote {\bibinfo {title} {Quantum simulation of
  Fermi-Hubbard models in semiconductor quantum-dot arrays},}\ }\href@noop {}
  {\bibfield  {journal} {\bibinfo  {journal} {Physical Review B}\ }\textbf
  {\bibinfo {volume} {78}},\ \bibinfo {pages} {075320} (\bibinfo {year}
  {2008})}\BibitemShut {NoStop}%
\bibitem [{\citenamefont {Radi\ifmmode~\acute{c}\else \'{c}\fi{}}\ \emph
  {et~al.}(2012)\citenamefont {Radi\ifmmode~\acute{c}\else \'{c}\fi{}},
  \citenamefont {Di~Ciolo}, \citenamefont {Sun},\ and\ \citenamefont
  {Galitski}}]{PhysRevLett.109.085303}%
  \BibitemOpen
  \bibfield  {author} {\bibinfo {author} {\bibfnamefont {J.}~\bibnamefont
  {Radi\ifmmode~\acute{c}\else \'{c}\fi{}}}, \bibinfo {author} {\bibfnamefont
  {A.}~\bibnamefont {Di~Ciolo}}, \bibinfo {author} {\bibfnamefont
  {K.}~\bibnamefont {Sun}}, \ and\ \bibinfo {author} {\bibfnamefont
  {V.}~\bibnamefont {Galitski}},\ }\bibfield  {title} {\enquote {\bibinfo
  {title} {Exotic quantum spin models in spin-orbit-coupled Mott insulators},}\
  }\href {\doibase 10.1103/PhysRevLett.109.085303} {\bibfield  {journal}
  {\bibinfo  {journal} {Phys. Rev. Lett.}\ }\textbf {\bibinfo {volume} {109}},\
  \bibinfo {pages} {085303} (\bibinfo {year} {2012})}\BibitemShut {NoStop}%
\bibitem [{\citenamefont {Jafari}\ \emph {et~al.}(2008)\citenamefont {Jafari},
  \citenamefont {Kargarian}, \citenamefont {Langari},\ and\ \citenamefont
  {Siahatgar}}]{jafari2008phase}%
  \BibitemOpen
  \bibfield  {author} {\bibinfo {author} {\bibfnamefont {Rouhollah}\
  \bibnamefont {Jafari}}, \bibinfo {author} {\bibfnamefont {M}~\bibnamefont
  {Kargarian}}, \bibinfo {author} {\bibfnamefont {A}~\bibnamefont {Langari}}, \
  and\ \bibinfo {author} {\bibfnamefont {M}~\bibnamefont {Siahatgar}},\
  }\bibfield  {title} {\enquote {\bibinfo {title} {Phase diagram and
  entanglement of the ising model with dzyaloshinskii-moriya interaction},}\
  }\href@noop {} {\bibfield  {journal} {\bibinfo  {journal} {Physical Review
  B}\ }\textbf {\bibinfo {volume} {78}},\ \bibinfo {pages} {214414} (\bibinfo
  {year} {2008})}\BibitemShut {NoStop}%
\bibitem [{\citenamefont {Hofstetter}\ \emph {et~al.}(2002)\citenamefont
  {Hofstetter}, \citenamefont {Cirac}, \citenamefont {Zoller}, \citenamefont
  {Demler},\ and\ \citenamefont {Lukin}}]{hofstetter2002high}%
  \BibitemOpen
  \bibfield  {author} {\bibinfo {author} {\bibfnamefont {Walter}\ \bibnamefont
  {Hofstetter}}, \bibinfo {author} {\bibfnamefont {J~Ignacio}\ \bibnamefont
  {Cirac}}, \bibinfo {author} {\bibfnamefont {Peter}\ \bibnamefont {Zoller}},
  \bibinfo {author} {\bibfnamefont {Eugene}\ \bibnamefont {Demler}}, \ and\
  \bibinfo {author} {\bibfnamefont {MD}~\bibnamefont {Lukin}},\ }\bibfield
  {title} {\enquote {\bibinfo {title} {High-temperature superfluidity of
  fermionic atoms in optical lattices},}\ }\href@noop {} {\bibfield  {journal}
  {\bibinfo  {journal} {Physical Review Letters}\ }\textbf {\bibinfo {volume}
  {89}},\ \bibinfo {pages} {220407} (\bibinfo {year} {2002})}\BibitemShut
  {NoStop}%
\bibitem [{\citenamefont {Maier}\ \emph {et~al.}(2005)\citenamefont {Maier},
  \citenamefont {Jarrell}, \citenamefont {Schulthess}, \citenamefont {Kent},\
  and\ \citenamefont {White}}]{maier2005systematic}%
  \BibitemOpen
  \bibfield  {author} {\bibinfo {author} {\bibfnamefont {Th~A}\ \bibnamefont
  {Maier}}, \bibinfo {author} {\bibfnamefont {M}~\bibnamefont {Jarrell}},
  \bibinfo {author} {\bibfnamefont {TC}~\bibnamefont {Schulthess}}, \bibinfo
  {author} {\bibfnamefont {PRC}\ \bibnamefont {Kent}}, \ and\ \bibinfo {author}
  {\bibfnamefont {JB}~\bibnamefont {White}},\ }\bibfield  {title} {\enquote
  {\bibinfo {title} {Systematic study of d-wave superconductivity in the 2d
  repulsive Hubbard model},}\ }\href@noop {} {\bibfield  {journal} {\bibinfo
  {journal} {Physical Review Letters}\ }\textbf {\bibinfo {volume} {95}},\
  \bibinfo {pages} {237001} (\bibinfo {year} {2005})}\BibitemShut {NoStop}%
\end{thebibliography}%

\end{document}